\documentclass[a4paper,12pt]{article}
\usepackage[cp1251]{inputenc}
\usepackage[T2A]{fontenc}
\usepackage[russian]{babel}
\usepackage[dvips]{graphicx}
\newcommand{\ga}{\alpha}

\newcommand{\go}{\omega}

\newcommand{\gl}{\lambda}

\newcommand{\gO}{\Omega}

\newcommand{\rf}[1]{(\ref{#1})}
\newcommand{\beq}[1]{\begin{equation} \label{#1}}
\newcommand{\eeq}{\end{equation}}
\newcommand{\beqn}{\begin{eqnarray*}}
\newcommand{\eeqn}{\end{eqnarray*}}
\newcommand{\beqa}[1]{\begin{eqnarray} \label{#1}}
\newcommand{\eeqa}{\end{eqnarray}}

\title{Long Time Oscillations of Wolf Number Series Autocorrelation Function and Possibility of Solar Activity Prediction}

\author{В.\,М.\,Журавлев\/\thanks{e-mail: zhvictorm@mail.ru},
С.\,В.\,Летуновский}

\begin{document}

\maketitle

\begin{abstract}
 Results of analysis of long time variations of solar activities based on monthly Wolf number series (1749-2011 years) are represented. Shown the presence of stable oscillations in autocorrelation function for the Wolf numbers series with a period near 42.5 years. The statistical model of variations of autocorrelation function based on assumption of nonstationarity process are constructed. The question of using the model to predict the solar activity is discussed.
\end{abstract}

В работах \cite{ZhL10,ZhL11} был рассмотрен вопрос о вероятностных характеристиках долгопериодической изменчивости солнечной активности на основе анализа ряда чисел Вольфа с 1749 по 2009 год \cite{NW}. Выбор ряда чисел Вольфа для анализа долгопериодических изменений с временными масштабами порядка 100 лет обусловлен тем, что этот набор данных является единственным достаточно достоверным свидетельством в форме количественных соотношений изменчивости солнечной активности на протяжении последних 250 лет. Основой исследований в этих работах являлся метод скользящих рядов длиной 100 лет с использованием метода моментов для оценивания параметров модельного распределения вероятностей чисел Вольфа. Основной вывод, который был сделан в указанных работах состоял в том, что в солнечной активности проявляются два различных механизма пятнообразовательной деятельности Солнца. Один представляет собой механизм формирования устойчивого вероятностного распределений чисел Вольфа типа экспоненциального распределения, аналогичного по форме распределению Больцмана. Второй же механизм можно охарактеризовать как механизм увеличения или снижения вероятности появления одновременно большого числа пятен, что можно рассматривать как проявление изменения вспышечной активности Солнца. Выявленные механизмы сопоставлялись с изменчивостью нитратов в льдах Антарктики и Гренладии, с которыми в среднем были выявлены определенные корреляции. Однако полученные в работах \cite{ZhL10,ZhL11} результаты не дают возможности применить построенные модели непосредственно к задаче построения прогноза солнечной активности, что является важным элементом современных исследований Солнца \cite{P10}. В настоящей работе исследуется методами аналогичными использованным в \cite{ZhL10,ZhL11} долгопериодическая изменчивость автокорреляционной функции ряда чисел Вольфа. Основной эффект, который обнаружен в результате этих исследований,  состоит в обнаружении устойчивых осцилляций автокорреляционной функции на отдельных отрезках времени длиной порядка 30 лет, на границе которых происходит резкий сдвиг фаз таких осцилляций без изменения периода осцилляций. На основе этих результатов предложена новая модель для случайного процесса, представляющего ряд чисел Вольфа. В работе рассмотрены основные свойства модели и проведено сравнение с наблюдаемыми значениями автокорреляционной функции ряда чисел Вольфа.

1. Оценка автокорреляционной функции для дискретного цифрового ряда наблюдений длиной $N$: $W_i,~i=1,\ldots,N$, используемая в данной работе, строится по стандартным правилам и имеет следующий вид:
  \beq{DefCor}
    R_k = \frac{1}{N-k-1}\sum\limits_{i=k}^N W_i W_{i-k}.
  \eeq
  Для полного ряда числе Вольфа (1749-2011) общий вид автокорреляционной функции представлен на рис. \rf{ris3ac0}.
  \begin{figure}
    \begin{center}
        \includegraphics[width=0.90\textwidth]{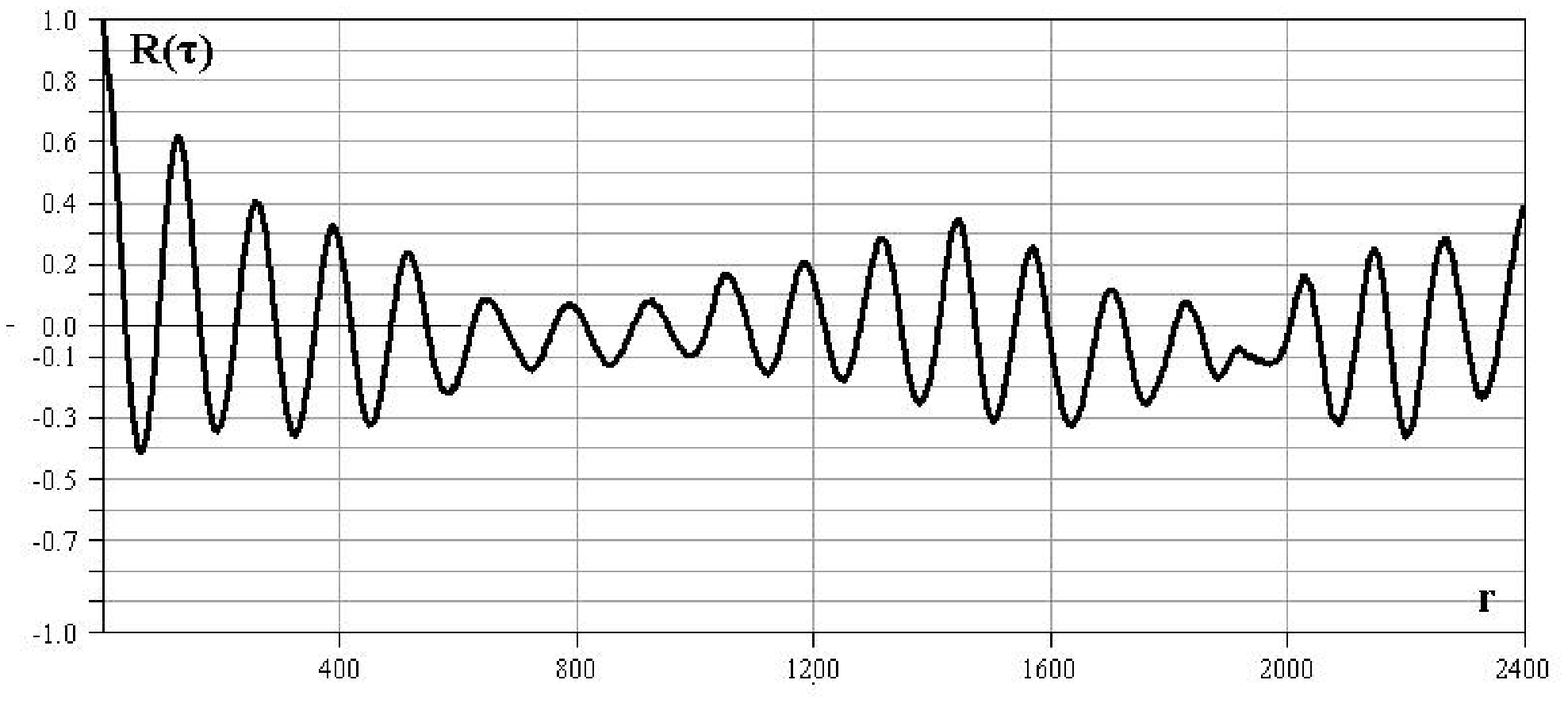}
    \end{center}
    \caption{Автокорреляционная функция полного ряда чисел Вольфа. Величина отсчёта 1 месяц, общая длина - 200 лет.}
    \label{ris3ac0}
    \end{figure}
  Наиболее важным свойством автокорреляционной функции, представленной на рис. \rf{ris3ac0} является не затухающий ее характер, что указывает на то, что процесс солнечной активности в форме чисел Вольфа является достаточно регулярным и сходным по свойствам со стационарными в широком смысле процессами \cite{Went}. Однако как было продемонстрировано в ряде работ по построению моделей предсказания, например, в работах \cite{ZhL10,ZhL11}, в реальности процесс  $W(t)$ является не стационарным. Именно это затрудняет построение прогностических моделей для солнечной активности

Метод скользящих рядов для анализа слабо нестационарных процессов состоит в проведении сравнительного анализа отдельных отрезков исходного ряда. Исходный ряд $X_{i},~~i=1,\ldots,N$ представляется в виде $M$ рядов $Y^{a}_i$ длиной $L$, сформированных с помощью следующего алгоритма:
$$
     Y^a_{i} = X_{i+s(a-1)},~~i=1,\ldots,L,~~a=1,\ldots,M.
$$
Длина исходного ряда $N$, длина сформированных рядов $L$, величина сдвига $S$ между рядами и число новых рядов $M$ связаны соотношением:
$$
      M= [(N-L)/S]+1,
$$
где $[x]$ - целая часть числа $x$. Сформированный из ряда ежемесячных чисел Вольфа набор из 479 рядов длиной $1200$ точек (100 лет) взятых через $S=4$ месяцев подвергался обработке для вычисления автокорреляционной функции каждого отдельного отрезка. Сводный набор из 159 автокорреляционных функций, взятых через $S=12$,  представлен на рис. 2
\begin{figure}
    \begin{center}
        \includegraphics[width=0.90\textwidth]{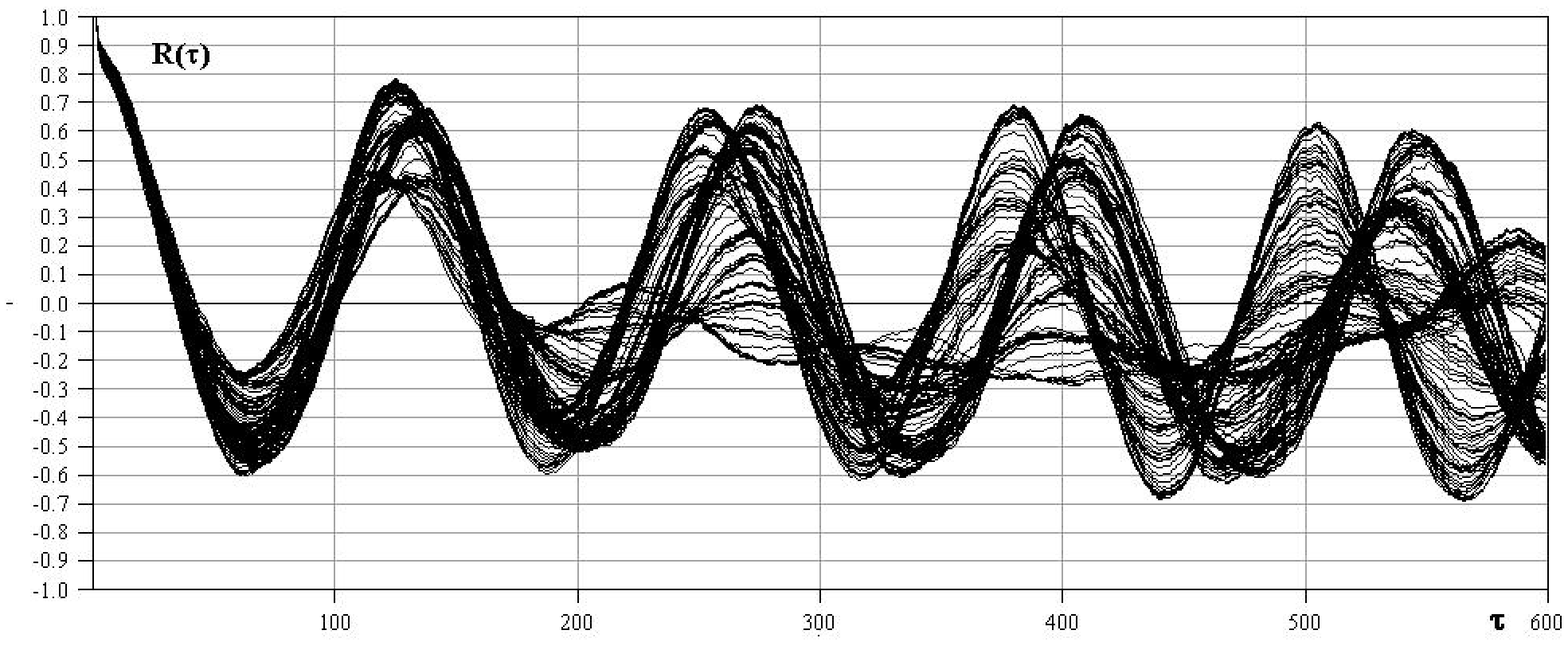}
    \end{center}
    \caption{Автокорреляционная функция отдельных отрезков ряда чисел Вольфа длиной 100 лет, взятых через 12 месяцев. Величина отсчёта по времени 1 месяц.}
    \label{ris3ac1}
    \end{figure}

Как показывает рис. 2 автокорреляционные функции группируются в отдельные кластеры, относящиеся к отдельным интервалам времени. В каждом кластере автокорреляционная функция меняется таким образом, что положение её нулей и экстремумов остаётся практически неизменным, а меняется лишь само значение функции в экстремумах. Но при переходе от одного временного кластера к другому положение нулей и экстремумов изменяется практически скачком.  В целом эти интервалы соотносятся с периодизацией долговременной изменчивости солнечной активности, полученной другими методами в работах \cite{ZhL10,ZhL11}. Это означает, что в те моменты времени, когда происходит резкое изменение вспышечной активности Солнца, скачком меняются и свойства самой автокорреляционной функции. Таким образом, выявленная в этих работах изменчивость подтверждается на новых данных. Однако особый интерес представляет то, как происходит изменчивость автокорреляционной функции в внутри отдельного временного кластера. Именно может дать полезную информацию для прогноза изменчивости солнечной активности.

2. Для анализа поведения автокорреляционной функции внутри выделенных временных кластеров её стабильности, использовался следующий алгоритм. Для удобства анализа и интерпретации были выбраны четыре отдельных кластера, содержащих автокорреляционные функции за период примерно 40 лет. При этом анализу подвергались автокорреляционные функции взятые со сдвигом в 4 месяца для надежности оценивания периодов их изменчивости.  В результате каждый кластер содержал по 120 отдельных автокорреляционных функции. Для каждого кластера вычислялись средние по кластеру автокорреляционные функции, которые затем вычитались из каждой функции кластера. Дальнейшему анализу подвергались остаточные ряды значений автокорреляционной функции с шагом между отсчётами в 1 месяц и длиной 50 лет (600 точек). Типичный вид содержимого кластера приведен на рис. 2.

\begin{figure}
    \begin{center}
        \includegraphics[width=0.90\textwidth]{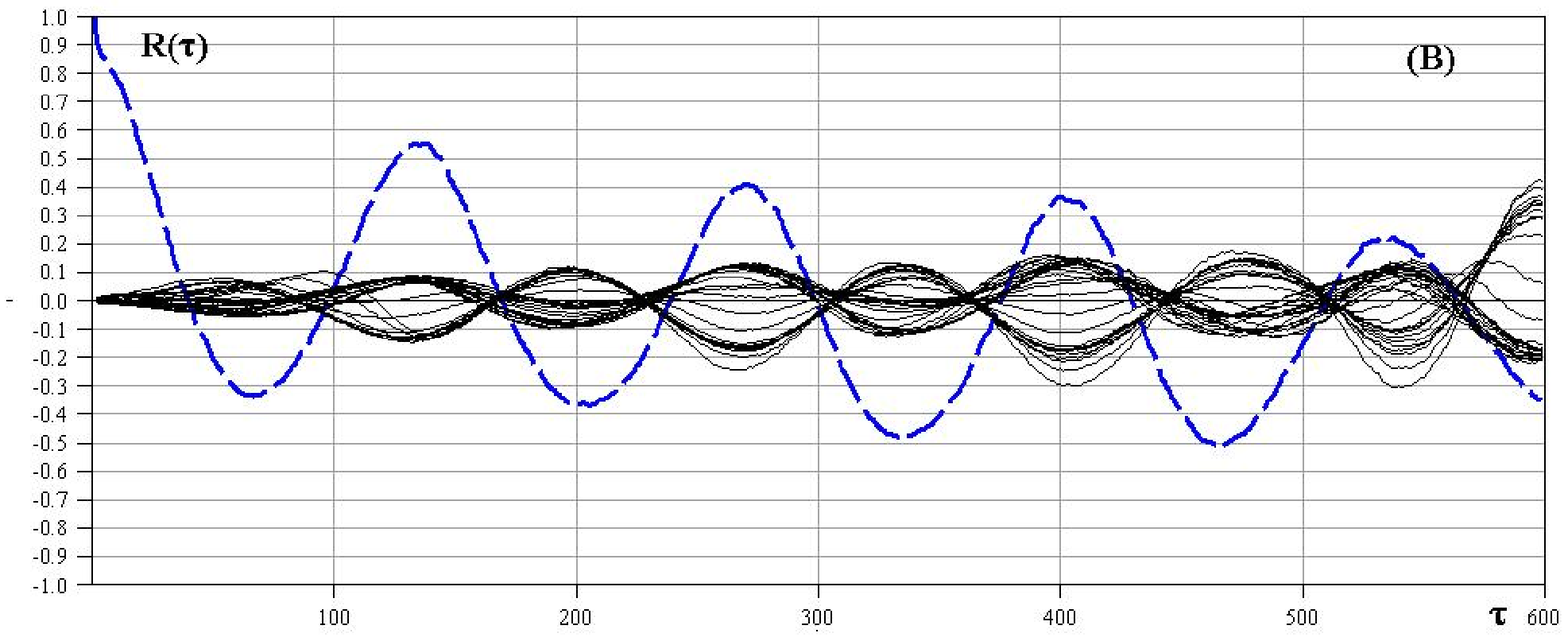}
    \end{center}
    \caption{Усредненная автокорреляционная функция $\overline{R_2}(\tau)$ второго кластера (пунктир) и остаточные ряды отклонений. Величина отсчёта 1 месяц.}
    \label{ris3ac2}
    \end{figure}

На графике \rf{ris3ac2} пунктирной линией приведена средняя по второму кластеру автокорреляционная функция, а тонкими сплошными линиями приведены графики остаточных функций. Как видно остаточные ряды совершают осцилляции вокруг нулевой функции таким образом, что положение узлов и пучностей этих осцилляций остаётся почти неизменным. Модель долговременных изменений автокорреляционной функции, соответствующую выявленным осцилляциям можно представить в виде разложения по эмпирическим ортогональным модам \cite{MEOM} на каждом из выделенных временных сегментов в следующем виде:
\beq{CorMod1}
    R^{(a)}(t,\tau) = R^{(a)}_0(\tau) + \sum\limits_{\ga=1}^L \sqrt{\gl^{(a)}_{\ga}} R^{(a)}_{\ga}(\tau) \xi^{(a)}_{\ga}(t).
\eeq
Здесь $a$ - номер временного сегмента устойчивых осцилляций, $R^{(a)}_0(\tau)$ - средняя функция на сегменте, $\gl^{(a)}_{\ga}$ - собственные числа мод с номером $\ga$ на сегменте с номером $a$, $R^{(a)}_{\ga}(\tau)$ - мода автокорреляционной функции, $\xi^{(a)}_{\ga}(t)$ -
изменение амплитуды моды со временем. Зависимость от  номера сегмента $a$ в этом разложении фактически означает наличие дополнительной изменчивости все функций от времени, которую можно охарактеризовать как некоторую ступенчатую функцию постоянную на сегменте, но скачком меняющую свое значение при переходе от одно сегмента к другому. Это поведение легко обнаруживается даже при визуальном анализе графиков на рис. \rf{ris3ac1} и \rf{ris3ac2}.
\begin{figure}
    \begin{center}
        \includegraphics[width=0.90\textwidth]{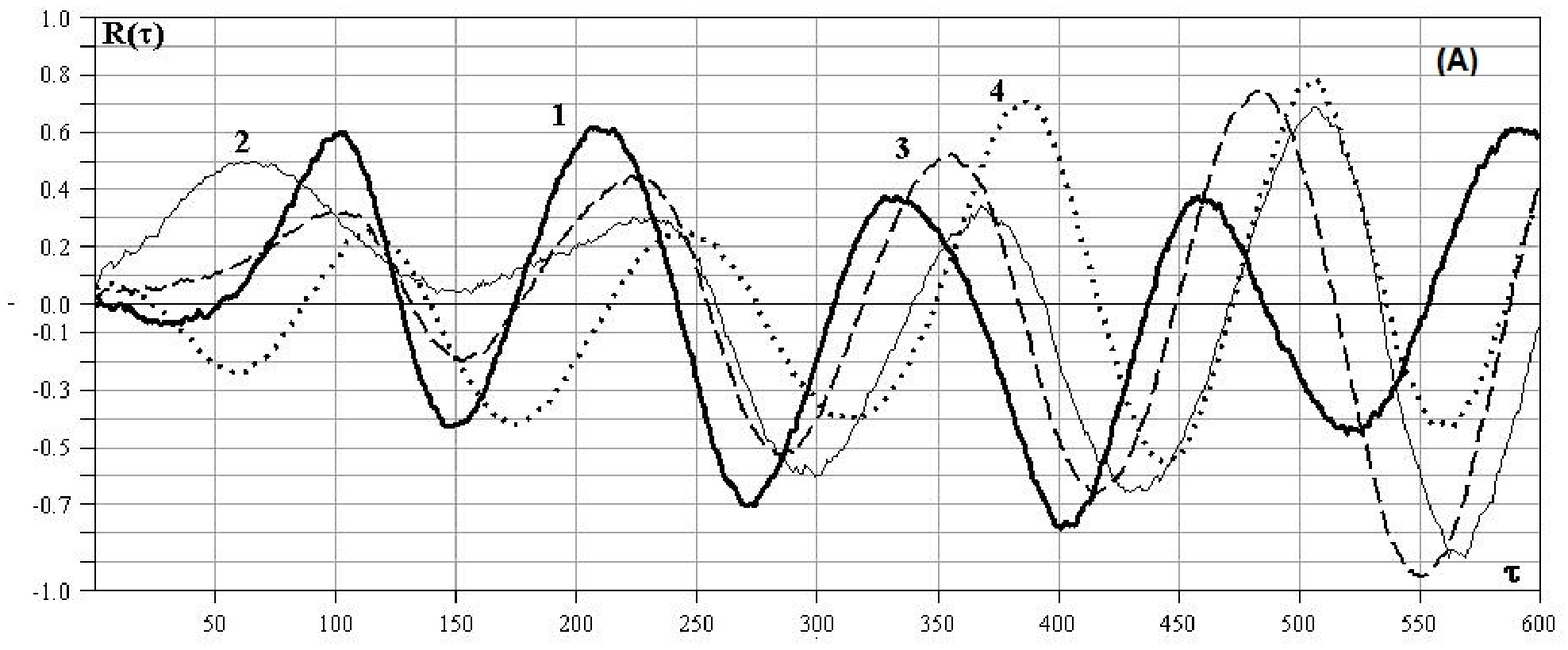}
        \includegraphics[width=0.90\textwidth]{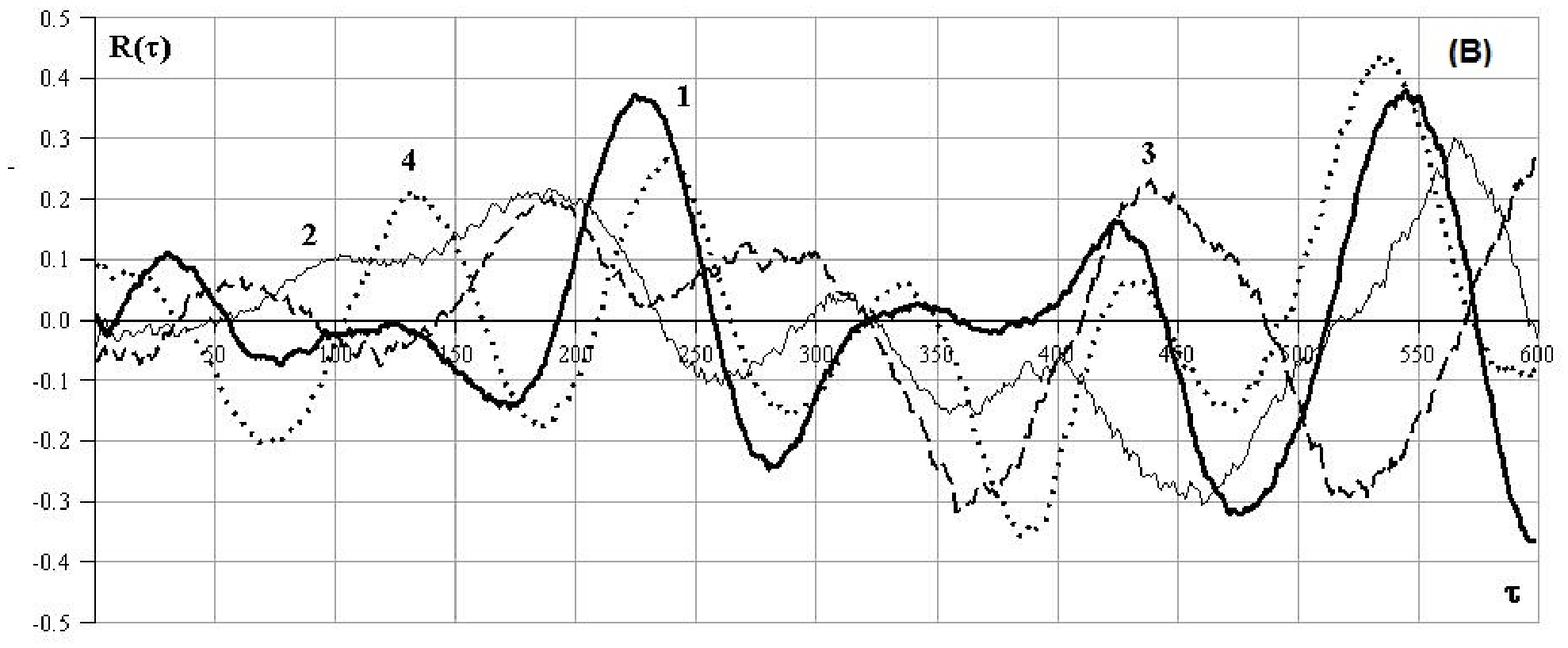}
        \includegraphics[width=0.90\textwidth]{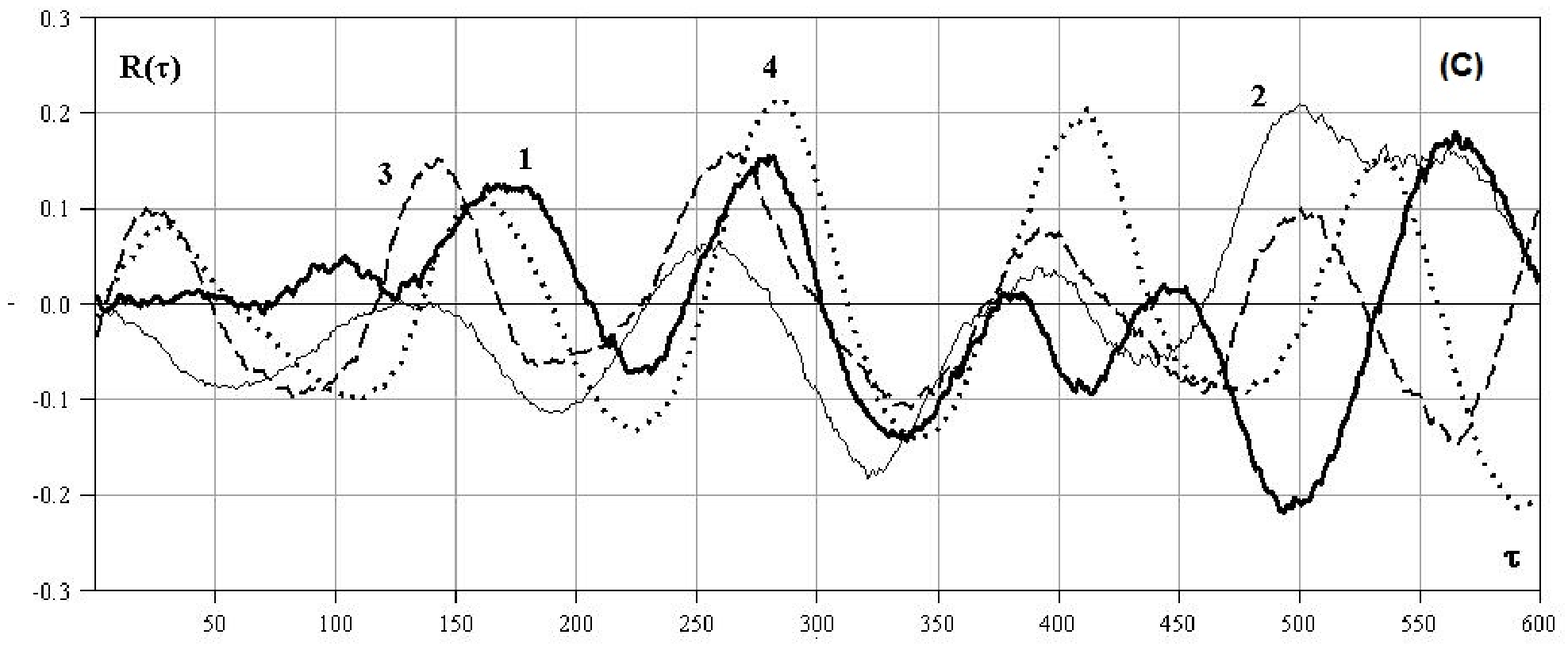}
    \end{center}
    \caption{Три первых моды (A-1,B-2,C-3) остаточных рядов автокорреляционной функции. 1,2,3,4 - номера сегментов}
    \label{ris3ac3}
    \end{figure}

3. Результаты анализа осцилляций внутри отдельных кластеров с помощью метода эмпирических ортогональных мод представлены на рис. \rf{ris3ac3} и \rf{ris3ac4}. На рис. 3 и рис. 4 приведены графики трёх первых ортогональных мод (рис. \rf{ris3ac3}) и графики их амплитуды (рис. \rf{ris3ac4}).
    \begin{figure}
    \begin{center}
        \includegraphics[width=0.90\textwidth]{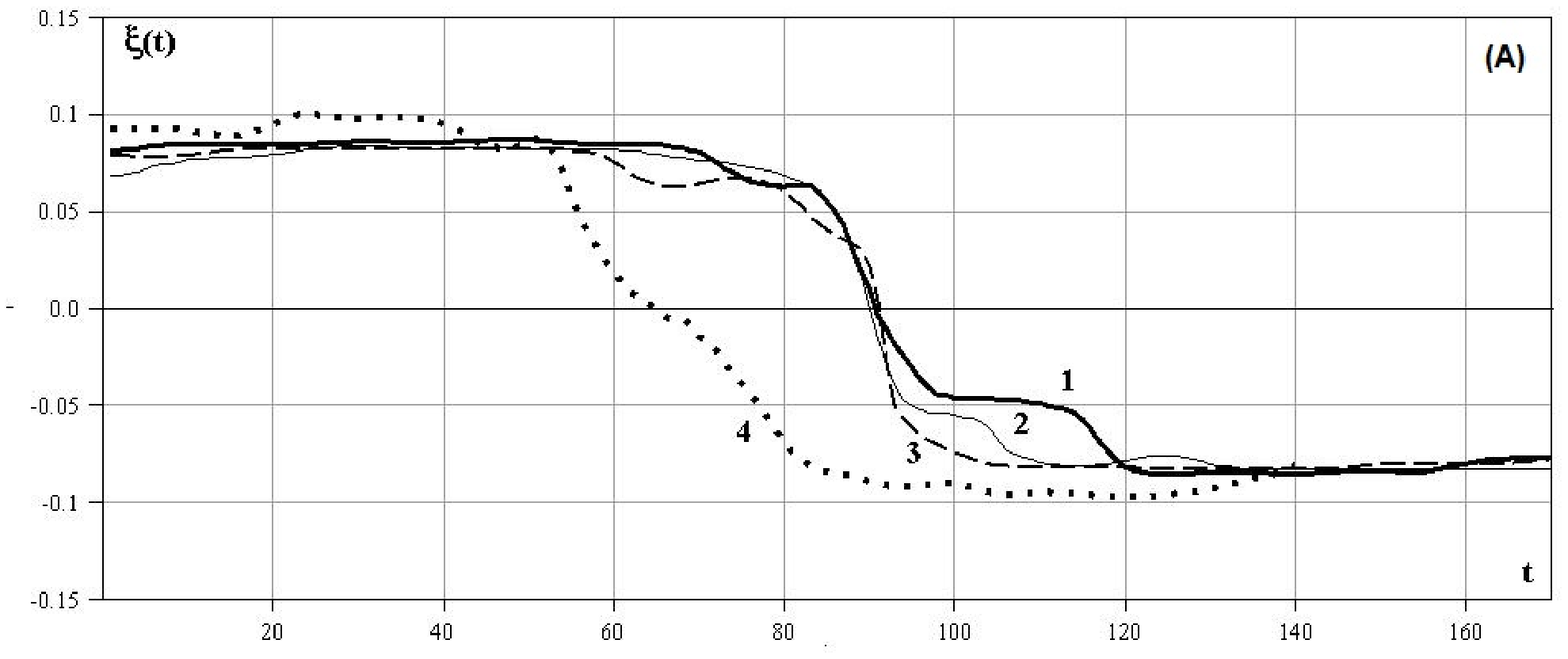}
        \includegraphics[width=0.90\textwidth]{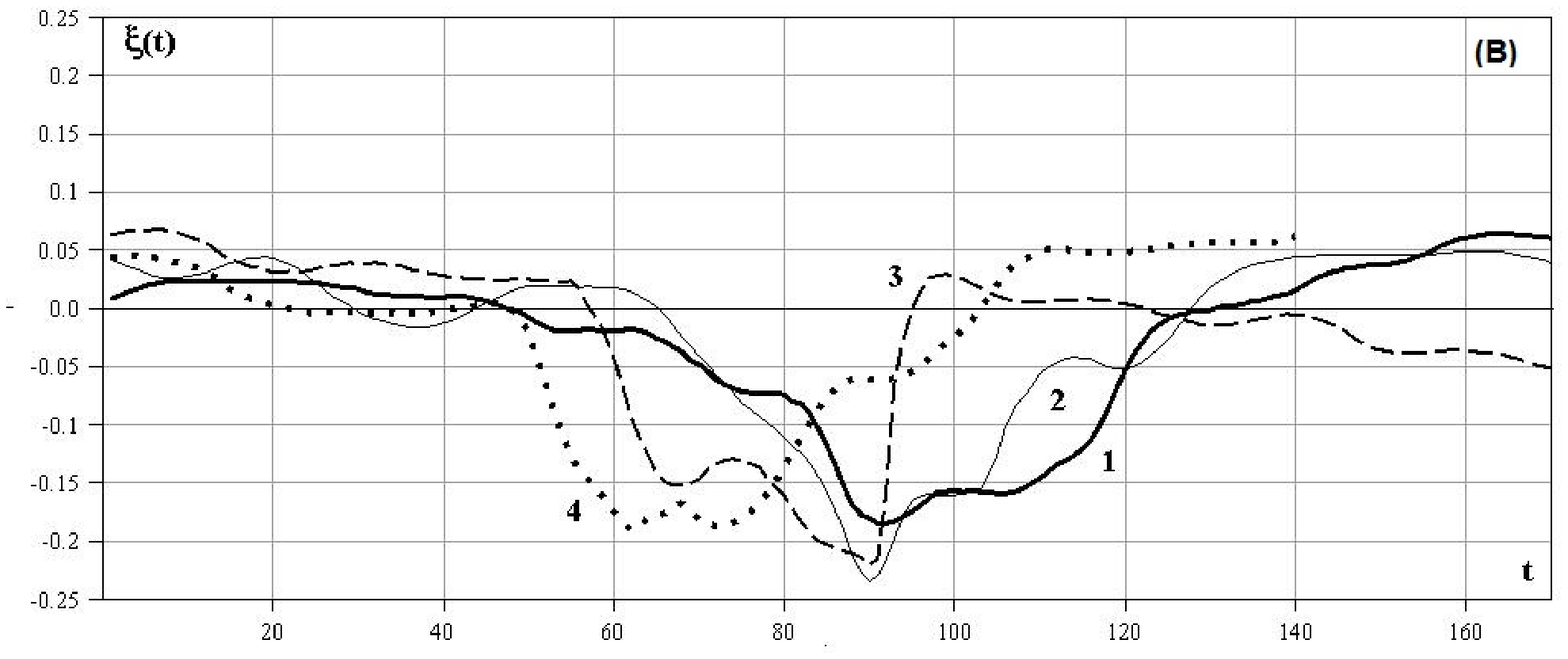}
        \includegraphics[width=0.90\textwidth]{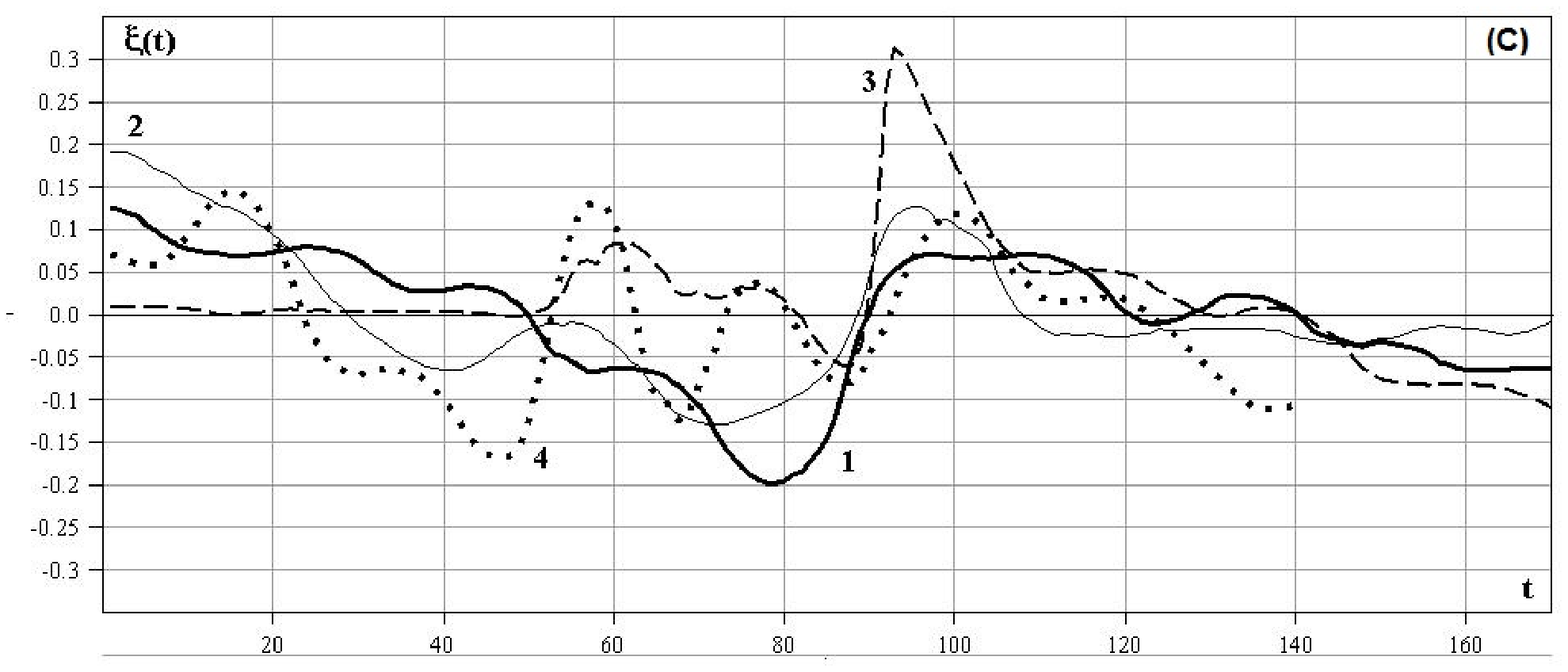}
    \end{center}
    \caption{Изменение амплитуды мод (A-1,B-2,C-3) со временем. 1,2,3,4 - номера сегментов}
    \label{ris3ac4}
    \end{figure}

Величина собственных чисел для первых трёх мод на всех четырёх сегментах представлена в таблице.

\begin{tabular}{|c|c|c|c|c|}
\hline
 N & \multicolumn{4}{c|}{Номер моды}\\
 \hline
 сегмента & $1$ & $2$ & $3$ & $4$ \\
 \hline
 1 & 277.0 & 124.3 &  66.2 & 45.0\\
\hline
 2 & 286.5 & 100.9 &  68.3 & 50.0\\
 \hline
 3 &  294.4 & 102.6 &  56.4 & 28.5\\
 \hline
 4 & 242.1 & 122.3 &  76.4 & 47.7\\
\hline
\end{tabular}

Рис. \rf{ris3ac4}(A) демонстрирует основной результат подбора длины сегментов. Этот результат состоит в том, что существует такая длина сегмента (170 месяцев или 42.5 года), что амплитуды первой основной моды ведут себя идентично на первых трёх сегментах. Длина четвёртого сегмента из-за недостаточно длины исходного ряда составляет 141 месяц или 31.25 лет. По всей видимости именно это обстоятельство приводит к отклонению поведения амплитуды первой моды на четвертом сегменте от поведения на остальных сегментах. Идентичность поведения первой моды на трёх первых сегментах означает, что амплитуда первой моды остаётся почти постоянной на первой половине сегмента, затем очень быстро в середине сегмента она очень быстро меняет свой знак на обратный и остаётся далее неизменной на второй половине сегмента. На границах сегментов происходит новая смена знака амплитуды первой моды за короткий промежуток времени. На границах же сегмента происходит скачкообразное изменение усреднённой по сегменту автокорреляционной функции. Таким образом, подобранная длина сегментов в 42.5 года (170 месяцев) соответствует в реальности основной частоте скачкообразных изменений амплитуды первой моды.

В целом аналогичное поведение наблюдается и у второй и третьей основных мод. Основные изменения амплитуды этих мод происходят так же в середине сегментов. Однако осцилляции, описываемые этими модами происходят с большей частотой. Если период осцилляций первой моды по характеру её поведения можно оценить как длину сегмента, то  периоды второй и третьей моды примерно кратны соответствующей основной частоте первой моды $\gO=2\pi/42.5$ [лет]${}^{-1}$ с учётом их скачкообразного изменения на границах сегмента. Для второй моды частота примерно равна $2\gO_0$, а для третьей $3\gO_0$.

Для более детального анализа поведения отдельных составляющих разложения \rf{CorMod1} при переходе от одного сегмента к другому рассмотрим отклонение функций $R_0^{(a)}(\tau)$ от общего среднего. Эти графики представлены на рис. \rf{ris3ac5}.

 \begin{figure}
    \begin{center}
        \includegraphics[width=0.90\textwidth]{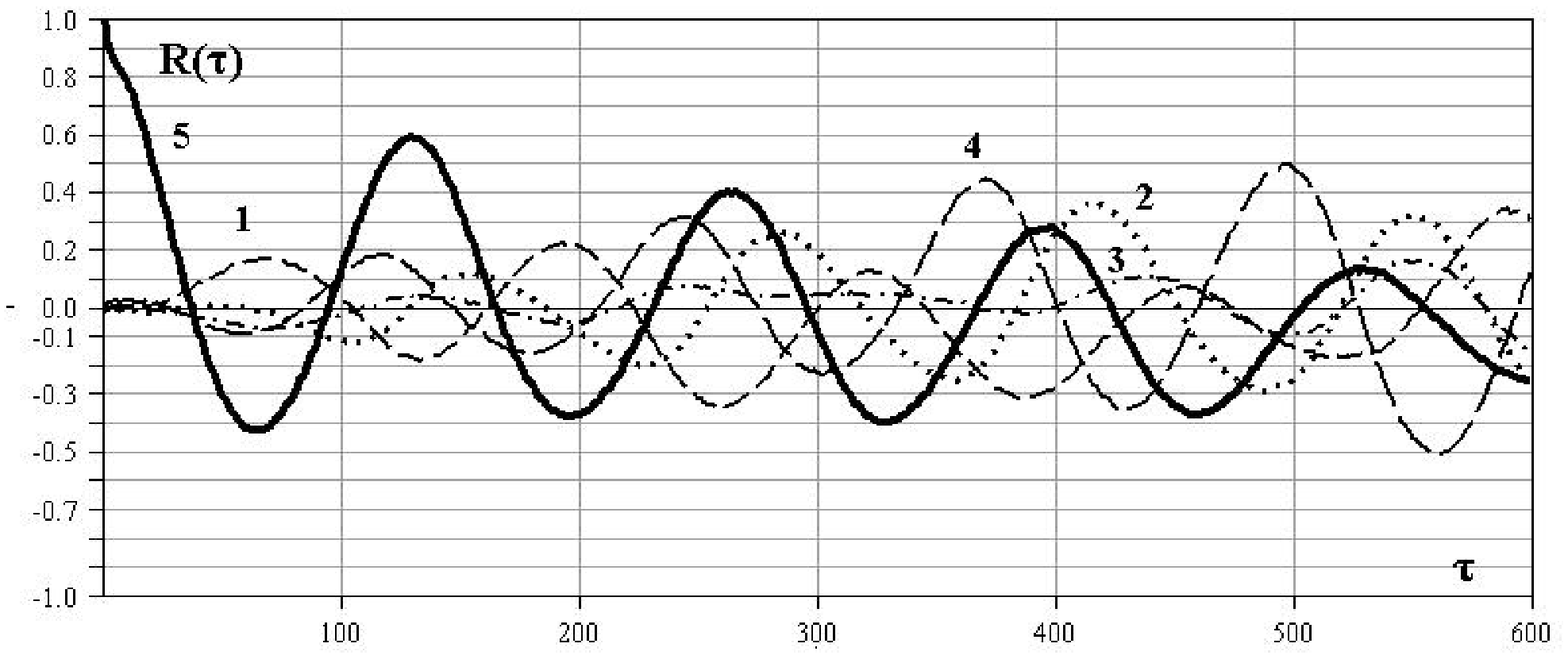}
    \end{center}
    \caption{Отклонение средних значений автокорреляционной функции по сегменту от средней по всем скользящим рядам. 1,2,3,4 - номера сегментов, 5 - средняя атокорреляционная функция по всем сдвигам скользящих рядов}
    \label{ris3ac5}
    \end{figure}

Как видно, функции $R^{(a)}_0(\tau)$ можно представить в следующем виде:
\beq{ACorMod}
      R^{(a)}_0(\tau)=R_0(\tau)+\rho^{(a)}(\tau).
\eeq
Функции $\rho^{(a)}(\tau)$ ведут примерно одинаково. Эти функции от параметра $\tau$ можно представить в виде одной гармоники, модулированной более низкочастотной гармоникой. Отличие функций   $\rho^{(a)}(\tau)$ друг от друга в основном сводится к сдвигу фаз высокочастотной гармоники при небольшом изменении формы низкочастотной огибающей. Фактически это означает, что частотный состав этих отклонений одинаков, а изменения со временем сводятся к изменению амплитуды коэффициентов гармонического разложения этих функций. Это наблюдение важно с точки зрения указания типа модели для такого поведения автокорреляционной функции.

4. Для того, что бы построить модель случайного процесса, описывающего наблюдаемые свойства автокорреляционной функции скользящих рядов чисел Вольфа, рассмотрим случайный процесс следующего вида:
\beq{DefModW}
    w(t) = \sum\limits_{k=1}^{\infty}A_k \cos(\go_k t + \phi_k),
\eeq
в котором амплитуды гармоник и их фазы являются случайными числами. Такого типа процесс при условии независимости случайных величин $A_k$ и $\phi_k$ и дополнительном требовании:
\beq{Condphi}
      \overline{\cos\phi_k} = \overline{\sin\phi_k} =0
\eeq
представляет собой стационарный в широком смысле процесс. Здесь черта сверху обозначает усреднение по ансамблю (вероятностному распределению). Стационарность в широком смысле означает:
$$
    \overline{w(t)w(t-\tau)} = R_w(\tau).
$$
Для процесса \rf{DefModW} имеем:
\beqn
     && R_w(\tau) = \sum\limits_{k=1}^{\infty}\sum\limits_{k=1}^{\infty}\overline{A_jA_k} \overline{\cos(\go_k t + \phi_k)\cos(\go_k (t-\tau) + \phi_j)}=\\
     && = \sum\limits_{k=1}^{\infty}\frac{\overline{A_k^2}}{2}\cos(\go_k\tau).
\eeqn

Однако, если для процесса \rf{DefModW} не предполагать независимости случайных величин $A_k$ и $\phi_k$, то этот процесс перестаёт быть процессом стационарным в широком смысле. Действительно, в этом случае имеем:
\beqn
    && R_w(\tau,t) = \sum\limits_{k=1}^{\infty}\frac{\overline{A_k^2}}{2}\cos(\go_k\tau)+\\
    && + \sum\limits_{k=1}^{\infty}\sum\limits_{j=1}^{\infty}\left\{ Q^+_{kj} \cos[(\go_k+\go_j)t-\go_j\tau ] + \right. \\
     && - P^+_{kj} \sin[(\go_k+\go_j)t-\go_j\tau ] + \\
    && + Q^-_{kj} \cos[(\go_k-\go_j)t+\go_j\tau ] + \\
    &&- \left. P^-_{kj} \sin[(\go_k-\go_j)t+\go_j\tau ]\right\}.
\eeqn
Здесь
\beqa{DefPQ}
      && Q^+_{kj} = \frac{\overline{A_kA_j\cos(\phi_k+\phi_j)}}{2} ,~
         P^+_{kj} = \frac{\overline{A_kA_j\sin(\phi_k+\phi_j)}}{2},\\ \nonumber
      && Q^-_{kj} = \frac{\overline{A_kA_j\cos(\phi_k-\phi_j)}}{2} ,~
         P^-_{kj} = \frac{\overline{A_kA_j\sin(\phi_k-\phi_j)}}{2}.
\eeqa
Полученное соотношение для $R_w(\tau,t)$ можно записать в более компактной форме:
\beqa{DefRttau}
    && R_w(\tau,t) = \sum\limits_{k=1}^{\infty}\frac{\overline{A_k^2}}{2}\cos(\go_k\tau)+\\
    \nonumber
    && + \sum\limits_{k=1}^{\infty}\Big(C_k(t)\cos\go_k\tau + S_k(t)\sin\go_k\tau\Big),
\eeqa
где
\beqn
   && C_j(t) = \sum\limits_{k=1}^{\infty} \Big[Q^+_{kj}\cos\gO^+_{kj}t-P^+_{kj}\sin\gO^+_{kj}t +\\
   && +Q^-_{kj}\cos\gO^-_{kj}t-P^-_{kj}\sin\gO^-_{kj}t \Big],\\
   && S_j(t) = \sum\limits_{k=1}^{\infty} \Big[Q^+_{kj}\sin\gO^+_{kj}t+P^+_{kj}\cos\gO^+_{kj}t +\\
   && +Q^-_{kj}\sin\gO^-_{kj}t+P^-_{kj}\cos\gO^-_{kj}t \Big],\\
   && \gO^-_{kj}=\go_k-\go_j,~~~\gO^+_{kj}=\go_k+\go_j.
\eeqn
Из этих соотношений видно, что процесс с коррелированными амплитудами и фазами является не стационарным процессом, причем его автокорреляционная функция имеет вид, аналогичный \rf{CorMod1}, который соответствует автокорреляционной функции ряда чисел Вольфа.

Сравнивая это представление с \rf{CorMod1} и \rf{ACorMod} приходим к выводу, что функцию $R_0(\tau)$ можно сопоставить первому слагаемому в \rf{DefRttau}:
$$
     R_0(\tau) =  \sum\limits_{k=1}^{\infty}\frac{\overline{A_k^2}}{2}\cos(\go_k\tau),
$$
которая описывает стационарную часть автокорреляционной функции.

5. Для усреднённого  ряда чисел Вольфа, который может служить основой прогноза изменчивости солнечной активности, из модели \rf{DefModW} можно получить следующее соотношение:
\beq{DefAW}
      W(t) = \overline{w(t)} = \sum\limits_{k=1}^{\infty}\Big( q_k\cos(\go_k t) + p_k\sin(\go_k t)\Big).
\eeq
где
$$
      q_k = \overline{A_k\cos\phi_k},~~p_k=-\overline{A_k\sin\phi_k}.
$$
Если предположить, что коррелируют между собой только амплитуды и фазы, соответствующие только одной частоте, а амплитуды и фазы других гармоник попарно статистически независимы между собой, то для вычисления величин $q_k,~p_k$, входящих в выражение \rf{DefAW}, можно воспользоваться следующими соотношениями:
\beqa{Eqpq}
    && Q^+_{kj} = \frac{p_kp_j-q_kq_j}{2},~~Q^-_{kj} = \frac{p_kp_j+q_kq_j}{2},\\ \nonumber
    && P^+_{kj} = \frac{p_kq_j+q_kp_j}{2},~~P^-_{kj} = \frac{p_kq_j-q_kp_j}{2}.
\eeqa
Однако при проведении гармонического анализа автокорреляционной функции с использованием этих соотношений задача о вычислении $q_k,~p_k$ будет некорректной в силу переопределённости системы уравнений \rf{Eqpq}. Поэтому для реализации такой процедуры необходимо разрабатывать специальную процедуру, регуляризирующую процесс вычислений чисел $q_k,~p_k$. Одним из способов регуляризации системы уравнений \rf{Eqpq} является предположение о кратности всех гармоник, входящих в представление \rf{DefModW}. Обнаруженные свойства разложения \rf{CorMod1} указывают на возможность применения такой процедуры для построения решения уравнений \rf{Eqpq}. При этом следует, конечно, учитывать, что утрачивается возможность описания существования    несоизмеримых гармоник в процессе представленном рядом чисел Вольфа. Другой вариант построения решения системы \rf{Eqpq} состоит в использовании метода наименьших квадратов.
Однако построение такого алгоритма выходит за рамки данной работы.

6. Выводы, которые можно сделать из проведённого анализа сводятся к следующему.
На основе анализа автокорреляционной функции скользящих рядов чисел Вольфа установлено, что солнечная активность остаётся относительно устойчивой на отрезках (сегментах) времени порядка 42.5 лет и испытывает на  границах этих отрезков скачкообразные изменения. При этом внутри каждого такого отрезка автокорреляционная функция испытывает осцилляции с частотами кратным частоте, соответствующей периоду 42.5 лет. Эти выводы в целом соответствуют результатам анализа ряда чисел Вольфа, полученным в работах \cite{ZhL10} и \cite{ZhL11}, и уточняют их.

Хотя сам ряд чисел Вольфа является нестационарным, однако незатухающий характер автокорреляционной функции полного ряда чисел Вольфа указывает на высокую регулярность самого процесса. Для описания такого поведения была предложена новая модель квазистационарного процесса, представляющего собой ряд Фурье со случайными амплитудами и фазами гармоник, случайные флуктуации которых коррелируют между собой (статистически зависимы). Как показано, такой процесс в целом отражает наблюдаемые свойства автокорреляционной функции скользящих рядов. Предложенная модель позволяет, анализируя данные о автокорреляционной функции скользящих рядов

Работа выполнена при поддержке РФФИ (11-01-00747-а, 10-01-00608-a) и проекта
    федеральной целевой программы "Научные и научно-педагогические
    кадры инновационной России" на 2009-2013 годы", проект НК-594П/8.

\end{document}